\documentclass[12pt]{article}%
\usepackage{amsfonts}
\usepackage{amsmath}
\usepackage{amssymb}
\usepackage{graphicx}%
\providecommand{\U}[1]{\protect\rule{.1in}{.1in}}
\newtheorem{theorem}{Theorem}
\newtheorem{acknowledgement}[theorem]{Acknowledgement}

\newtheorem{axiom}[theorem]{Axiom}

\newtheorem{conjecture}[theorem]{Conjecture}

\newtheorem{definition}[theorem]{Definition}

\newtheorem{proposition}[theorem]{Proposition}

\newenvironment{proof}[1][Proof]{\noindent\textbf{#1.} }{\ \rule{0.5em}{0.5em}}
\begin{document}

\title{Quantum Blobs}
\author{Maurice A. de Gosson\thanks{maurice.de.gosson@univie.ac.at}\\\ \textit{Universit\"{a}t Wien, }\\\textit{Fakult\"{a}t f\"{u}r Mathematik }\\\textit{NuHAG}, \textit{A-1090 Wien}}
\maketitle

\begin{abstract}
Quantum blobs are the smallest phase space units of phase space compatible
with the uncertainty principle of quantum mechanics and having the symplectic
group as group of symmetries. Quantum blobs are in a bijective correspondence
with the squeezed coherent states from standard quantum mechanics, of which
they are a phase space picture. This allows us to propose a substitute for
phase space in quantum mechanics. We study the relationship between quantum
blobs with a certain class of level sets defined by Fermi for the purpose of
representing geometrically quantum states.

\end{abstract}

\begin{center}
\textbf{To Basil Hiley, Physicist and Mathematician, on his 75th birthday}
\end{center}

AMS classification 2010: 81S30; 81S10; 81Q65;  

\section*{Introduction}

\subsection*{Basil Hiley and me}

It is indeed an honour and a pleasure to Contribute to Basil Hiley's
Festschrift. When I met Basil for the first time (it was in the late 90s,
during my Swedish exile) I was immediately fascinated not only by his vision
of quantum mechanics and its philosophy, but also by the man himself; I was
immediately charmed by his utterly unassuming and gentlemanly manners together
with his British humour. Basil patiently explained to me the subtleties of the
causal interpretation of quantum mechanics\ and of the Implicate Order; our
conversations were invariably accompanied by a cup of strong Assam tea, his
favourite beverage (during daytime, that is; later at night we occasionally
replaced the cup of tea by a glass of a beverage known in France as
\textit{Pastis}). Of course, I already had read a lot about the causal
interpretation of quantum mechanics, but my knowledge and understanding of
this theory was merely on an abstract mathematical level. Thanks to Basil's
pedagogical skills \textit{Physics} now entered the scene and helped me to
understand some of the deep implications of the causal interpretation.
However, Basil also was a patient and empathetic listener, always eager to
hear about new developments in mathematics (Basil is not only a brilliant
physicist, he also has an excellent taste for mathematics). When I explained
to him my ideas on the uncertainty principle and introduced him to the
\textquotedblleft symplectic camel\textquotedblright\ and \textquotedblleft
quantum blobs\textquotedblright, he immediately became very enthusiastic and
encouraged me to pursue the approach I had initiated in some recent papers. He
was even kind enough to honour me by writing a foreword to my book
\cite{principi} where I explained some of these ideas. Therefore I could do no
less than to write this modest contribution to the \textquotedblleft Hiley
Festschrift\textquotedblright\ as a tribute to my friend Basil for his
Helsinki birthday party!

\subsection*{Contents}

In this paper I establish a fundamental correspondence of a geometric nature
between the squeezed coherent states familiar from quantum optics, and quantum
blobs. The latter are related\ the principle of the symplectic camel, which is
a deep topological property of canonical transformations, and allow a
\textquotedblleft coarse graining\textquotedblright\ of phase space in units
which are symplectic deformations of phase space balls with radius
$\sqrt{\hbar}$.

This paper is structured as follows. I begin by reviewing in Sect. 1 the main
definitions and properties of squeezed coherent states; In Sect. 2 I introduce
the notion of quantum blob which we discuss from a purely geometric point of
view. In Sect. 3 the fundamental correspondence between squeezed coherent
states and quantum blobs is established; this correspondence which is denoted
by $\mathcal{G}$ is bijective (that is one-to-one and onto); its definition is
made possible using the theory of the Wigner transform of Gaussian functions.
In Sect. 4 I prove the fundamental statistical property of quantum blobs: they
are a geometric picture of minimum uncertainty. Finally, in Sect. 5, I shortly
discuss the relationship between quantum blobs and a certain level set
introduced in 1930 by Enrico Fermi and which seems to have been almost
unnoticed in the Scientific literature. The paper ends with some conjectures
and a discussion of related topics I plan to develop in further work.

\subsection*{Notation}

The phase space $\mathbb{R}_{z}^{2n}\equiv\mathbb{R}_{x}^{n}\oplus
\mathbb{R}_{p}^{n}$ ($n\geq1$) is equipped with the standard symplectic form
$\sigma(z,z^{\prime})=p\cdot x^{\prime}-p^{\prime}\cdot x$ if $z=(x,p)$,
$z^{\prime}=(x^{\prime},p^{\prime})$. We are writing $x=(x_{1},...,x_{n})$,
$p=(p_{1},...,p_{n})$, and $p\cdot x=p_{1}x_{1}+\cdot\cdot\cdot+p_{n}x_{n}$ is
the usual Euclidean scalar product of $p$ and $x$. Equivalently $\sigma
(z,z^{\prime})=Jz\cdot z^{\prime}$ where $J=%
\begin{pmatrix}
0_{n\times n} & I_{n\times n}\\
-I_{n\times n} & 0_{n\times n}%
\end{pmatrix}
$ is the standard symplectic matrix. The group of linear automorphisms of
$\mathbb{R}_{z}^{2n}$ is denoted by $\operatorname*{Sp}(2n,\mathbb{R})$ and
called the standard symplectic group. We have $S\in\operatorname*{Sp}%
(2n,\mathbb{R})$ if and only if $S$ is a linear mapping $\mathbb{R}_{z}%
^{2n}\longrightarrow\mathbb{R}_{z}^{2n}$ such that $S^{T}JS=J$.

\begin{acknowledgement}
This work has been supported by the Austrian Research Agency FWF
(Projektnummer P20442-N13).
\end{acknowledgement}

\section{Squeezed Coherent States}

For details and complements see the seminal paper by Littlejohn
\cite{Littlejohn}; Folland \cite{fo89} also contains valuable information.

The archetypical example is that of the fiducial (or standard, or vacuum)
coherent state
\begin{equation}
\Phi^{\hbar}(x)=(\pi\hbar)^{-n/4}e^{-|x|^{2}/2\hbar} \label{fid}%
\end{equation}
where the factor $(\pi\hbar)^{-n/4}$ is introduced in order to ensure
normalization. It was first systematically used by Schr\"{o}dinger in 1926.
The notation $\Phi^{\hbar}(x)=\langle x|0\rangle$ is also widely used in
quantum mechanics. It represents the ground state of the isotropic harmonic
oscillator; alternatively it is an eigenstate of the annihilation operator
with eigenvalue zero. More generally one wants to consider Gaussians of the
type
\begin{equation}
\Phi_{X,Y}^{\hbar}(x)=(\pi\hbar)^{-n/4}(\det X)^{1/4}e^{-\frac{1}{2\hbar
}(X+iY)x\cdot x} \label{Gauss1}%
\end{equation}
where $X$ and $Y$ are real symmetric $n\times n$ matrices, $X$ positive
definite; we have $||\Phi_{X,Y}^{\hbar}||_{L^{2}}=1$. The Gaussian
(\ref{Gauss1}) is called a squeezed (or generalized) coherent state. Let
$\widehat{T}^{\hbar}(z_{0})$ be the Heisenberg--Weyl operator defined for a
function $\Psi\in L^{2}(\mathbb{R}^{n})$ by
\begin{equation}
\widehat{T}^{\hbar}(z_{0})\psi(x)=e^{\frac{i}{\hbar}(p_{0}\cdot x-\tfrac{1}%
{2}p_{0}\cdot x_{0})}\psi(x-x_{0}). \label{heisenberg}%
\end{equation}
We can still go one step further and define the shifted squeezed coherent
state
\begin{equation}
\Phi_{X,Y,z_{0}}^{\hbar}(x)=\widehat{T}^{\hbar}(z_{0})\Phi_{X,Y}^{\hbar}(x)
\label{gauzo}%
\end{equation}
where $\widehat{T}^{\hbar}(z_{0})$ is the Heisenberg--Weyl operator: if $\Psi$
is a function on configuration space $\mathbb{R}_{x}^{n}$ then
\begin{equation}
\widehat{T}^{\hbar}(z_{0})\Psi(x)=e^{\frac{i}{\hbar}(p_{0}\cdot x-\frac{1}%
{2}p_{0}\cdot x_{0})}\Psi(x-x_{0}). \label{trans}%
\end{equation}
We will write from now on
\begin{equation}
M=X+iY\text{ \ , \ }\Phi_{M}^{\hbar}=\Phi_{X,Y}^{\hbar}\text{ \ , \ }%
\Phi_{M,z_{0}}^{\hbar}=\widehat{T}^{\hbar}(z_{0})\Phi_{M}^{\hbar}.
\label{nota}%
\end{equation}

The important thing is that squeezed coherent states are naturally obtained
from the fiducial state (\ref{fid}) by letting metaplectic operators act on
it. Let us explain this property shortly; for details see for instance de
Gosson \cite{Birk}. The symplectic group $\operatorname*{Sp}(2n,\mathbb{R})$
has a covering group of order two, the metaplectic group $\operatorname*{Mp}%
(2n,\mathbb{R})$. That group consists of unitary operators (the metaplectic
operators) acting on $L^{2}(\mathbb{R}^{n})$. There are several equivalent
ways to describe the metaplectic operators. For our purposes the most
tractable is the following: assume that $S\in\operatorname*{Sp}(2n,\mathbb{R}%
)$ has the block-matrix form%
\begin{equation}
S=%
\begin{pmatrix}
A & B\\
C & D
\end{pmatrix}
\text{ \ with \ }\det B\neq0. \label{free}%
\end{equation}
The condition $\det B\neq0$ is not very restrictive, because one shows (de
Gosson \cite{principi,Birk}) that every $S\in\operatorname*{Sp}(2n,\mathbb{R}%
)$ can be written (non uniquely however) as the product of two symplectic
matrices of the type above; moreover the symplectic matrices arising as
Jacobian matrices of Hamiltonian flows determined by physical Hamiltonians of
the type \textquotedblleft kinetic energy plus potential\textquotedblright%
\ are of this type for almost every time $t$. To the matrix (\ref{free}) we
associate the following quantities:

\begin{itemize}
\item A quadratic form
\[
W(x,x^{\prime})=\frac{1}{2}DB^{-1}x\cdot x-B^{-1}x\cdot x^{\prime}+\frac{1}%
{2}B^{-1}Ax^{\prime}\cdot x^{\prime}%
\]
defined on the double configuration space $\mathbb{R}_{x}^{n}\times
\mathbb{R}_{x}^{n}$; the matrices $DB^{-1}$ and $B^{-1}A$ are symmetric
because $S$ is symplectic ($W(x,x^{\prime})$ is often called \textquotedblleft
Hamilton's characteristic function\textquotedblright\ Goldstein
\cite{HGoldstein}) in mechanics, or \textquotedblleft
eikonal\textquotedblright\ in optics; it is closely related to the notion of
action de Gosson \cite{principi,Birk});

\item The complex number $\Delta(W)=i^{m}\sqrt{|\det B^{-1}|}$ where $m$
(\textquotedblleft Maslov index\textquotedblright) is chosen in the following
way: $m=0$ or $2$ if $\det B^{-1}>0$ and $m=1$ or $3$ if $\det B^{-1}<0$.
\end{itemize}

The two metaplectic operators associated to $S$ are then given by%
\begin{equation}
\widehat{S}\Psi(x)=\left(  \tfrac{1}{2\pi i}\right)  ^{n/2}\Delta(W)\int
e^{\frac{i}{\hbar}W(x,x^{\prime})}\Psi(x^{\prime})d^{n}x^{\prime}.
\label{meta}%
\end{equation}
The fact that we have two possible choices for the Maslov index shows that the
metaplectic operators occur in pairs $\pm\widehat{S}$; this of course is just
a reflection of the fact that $\operatorname*{Mp}(2n,\mathbb{R})$ is a
two-fold covering group of $\operatorname*{Sp}(2n,\mathbb{R})$.

The action of $\operatorname*{Mp}(2n,\mathbb{R})$ on squeezed coherent states
is given by the following result:

\begin{proposition}
\label{prop1}Let $\widehat{S}\in\operatorname*{Mp}(2n,\mathbb{R})$ be one of
the two metaplectic operators corresponding to the symplectic matrix $S=%
\begin{pmatrix}
A & B\\
C & D
\end{pmatrix}
$ (we do not make the assumption $\det B\neq0$). Then
\begin{equation}
\widehat{S}\Phi_{M}^{\hbar}=e^{\frac{i}{\hbar}\gamma(\widehat{S})}\Phi_{M_{S}%
}^{\hbar}\text{ \ with \ }M_{S}=i(AM+iB)(CM+iD)^{-1} \label{sc1}%
\end{equation}
where $e^{\frac{i}{\hbar}\gamma(\widehat{S})}$ is a phase factor such that
$\gamma(-\widehat{S})=\gamma(\widehat{S})+i\pi\hbar$ [ the matrix $CM+iD$ is
never singular]. More generally we have:%
\begin{equation}
\widehat{S}\Phi_{M,z_{0}}^{\hbar}=e^{\frac{i}{\hbar}\gamma(\widehat{S})}%
\Phi_{M_{S},Sz_{0}}^{\hbar}. \label{sc2}%
\end{equation}

\end{proposition}

\begin{proof}
See Folland \cite{fo89}, Littlejohn \cite{Littlejohn}.
\end{proof}

This important result motivates the following definition:

\begin{definition}
The set $\mathcal{CS(}n,\mathbb{R})$ of all squeezed coherent states consists
of all $\{e^{\frac{i}{\hbar}\gamma}\Phi_{M,z_{0}}^{\hbar}\}$ where $\gamma$ is
an arbitrary real phase.
\end{definition}

We thus do not distinguish between $e^{\frac{i}{\hbar}\gamma}\Phi_{M,z_{0}%
}^{\hbar}$ and $e^{\frac{i}{\hbar}\gamma^{\prime}}\Phi_{M,z_{0}}^{\hbar}$; we
will often omit the prefactor $e^{\frac{i}{\hbar}\gamma}$. Proposition
\ref{prop1} can now be restated in terms of a group action:
\begin{gather*}
\operatorname*{Mp}(2n,\mathbb{R})\times\mathcal{CS(}n,\mathbb{R}%
)\longrightarrow\mathcal{CS(}n,\mathbb{R})\\
(\widehat{S},e^{\frac{i}{\hbar}\gamma}\Phi_{M}^{\hbar})\longmapsto e^{\frac
{i}{\hbar}(\gamma+\gamma(\widehat{S}))}\Phi_{M_{S}}^{\hbar}.
\end{gather*}

We will come back to this action in a moment and give a geometric picture of
it in terms of phase space ellipsoids.

An important property of Proposition \ref{prop1} above is that $\mathcal{CS(}%
n,\mathbb{R})$ is preserved by Hamiltonian flows arising from quadratic
Hamiltonian functions, i.e. Hamiltonians of the general type%
\begin{equation}
H(z)=\frac{1}{2}Rz\cdot z \label{quad}%
\end{equation}
where $R$ is a real symmetric matrix. When $H$ is of the physical type
\textquotedblleft kinetic energy plus potential\textquotedblright\ this
amounts considering potentials which are quadratic forms $\frac{1}{2}\Omega
x\cdot x$ in the position variables (generalized harmonic oscillator):%
\begin{equation}
H(z)=\frac{1}{2m}|p|^{2}+\frac{1}{2}\Omega x\cdot x. \label{home}%
\end{equation}

For Hamiltonians of the type (\ref{quad}) the flow determined by the Hamilton
equations%
\begin{equation}
\dot{x}=\nabla_{p}H(x,p)\text{ \ , \ }\dot{p}=-\nabla_{x}H(x,p) \label{ham0}%
\end{equation}
consists of linear canonical transformations (Arnol'd \cite{ar89}, Goldstein
\cite{HGoldstein}, de Gosson \cite{Birk}). In fact, rewriting these equations
in the form $\dot{z}=JXz$ with $X=-JR$ the explicit solution is given by
$z_{t}=(x_{t},p_{t})=e^{tX}z_{0}$. The matrix $X$ belongs to the symplectic
Lie algebra $\mathfrak{sp}(2n,\mathbb{R})$ (because $XJ+JX^{T}=0$, see Folland
\cite{fo89} or de Gosson \cite{principi,Birk}) hence the matrices
$S_{t}=e^{tX}$ are symplectic. For instance, for the generalized harmonic
oscillator (\ref{home}) the Hamilton equations are $\dot{x}=p/m$ and $\dot
{p}=-\Omega x$ and we have $X=%
\begin{pmatrix}
0 & 1/m\\
-\Omega & 0
\end{pmatrix}
$.

It follows from the theory of the metaplectic group that together with the
theory of covering spaces (see e.g. Folland \cite{fo89}, de Gosson
\cite{Birk}) that to the path $t\longmapsto S_{t}=e^{tX}$ of symplectic
matrices corresponds a unique path $t\longmapsto\widehat{S}_{t}$ of
metaplectic operators such that $\widehat{S}_{0}$ is the identity. The
remarkable fact is that this family of operators $\widehat{S}_{t}$ is just
precisely the quantum flow determined by Schr\"{o}dinger's equation%
\begin{equation}
i\hbar\frac{\partial\Psi}{\partial t}=H(x,-i\hbar\nabla_{x})\Psi\label{schrh}%
\end{equation}
where $H(x,-i\hbar\nabla_{x})$ is the (Weyl) quantization of the quadratic
Hamiltonian (\ref{quad}); for instance when $H$ has the physical type
(\ref{home}) this equation is just the usual equation%
\begin{equation}
i\hbar\frac{\partial\Psi}{\partial t}=\left[  -\frac{\hbar^{2}}{2m}\nabla
_{x}^{2}+\frac{1}{2}\Omega x\cdot x\right]  \Psi. \label{schr3}%
\end{equation}
Thus the solution of (\ref{schrh}) is given by the simple formula%
\begin{equation}
\Psi(x,t)=\widehat{S}_{t}\Psi_{0}(x)\text{ \ , \ }\Psi_{0}(x)=\Psi(x,0)
\label{solpsi}%
\end{equation}
In particular, if the initial wavefunction $\Psi_{0}(x)$ is a coherent state
$\Phi_{M_{0},z_{0}}^{\hbar}$ Proposition \ref{prop1}) shows that the solution
$\Psi(x,t)$ is explicitly given by
\begin{equation}
\Psi(x,t)=e^{\frac{i}{\hbar}\mathcal{\gamma}(t)}\Phi_{M_{t},z_{t}}^{\hbar}(x)
\label{sol1}%
\end{equation}

\begin{itemize}
\item $z_{t}=(x_{t},p_{t})$ is the solution of Hamilton's equations $\dot
{x}=\nabla_{x}H$, $\dot{p}=-\nabla_{p}H$ passing through the point $z_{0}$ at
time $t=0$;

\item $M_{t}$ is calculated using formula (\ref{sc1}): write $S_{t}$ as a
symplectic block matrix $%
\begin{pmatrix}
A_{t} & B_{t}\\
C_{t} & D_{t}%
\end{pmatrix}
$; then%
\begin{equation}
M_{t}=i(A_{t}M_{0}+iB_{t})(C_{t}M_{0}+iD_{t})^{-1}; \label{mabcd}%
\end{equation}

\end{itemize}

One proves (see for instance Nazaikiinskii et al. \cite{Naza}) that

\begin{itemize}
\item The phase $\mathcal{\gamma}(t)$, is the symmetrized action integral
\begin{equation}
\mathcal{\gamma}(t)=\int_{0}^{t}\left(  \tfrac{1}{2}\sigma(z_{\tau},\dot
{z}_{\tau})-H\right)  d\tau. \label{action}%
\end{equation}

\end{itemize}

\section{Quantum Blobs}

Quantum blobs are minimum uncertainty units which are measured using not
volume, but rather symplectic capacity, which has the properties of an area
--that is of action! Besides the fact that they allow a geometric description
of the uncertainty principle \cite{de02-2,de03-2,de04,de05,Birk} (of which the
reader will find a precise description in next subsection), we are going to
see that they are intimately related to the notion of squeezed coherent
states, of which it can be considered as a phase space geometric picture.

By definition, a quantum blob is a subset $\mathcal{QB}^{2n}=\mathcal{QB}%
^{2n}(z_{0},S)$ of $\mathbb{R}_{z}^{2n}$ which can be deformed into the phase
space ball $B^{2n}(\sqrt{\hbar}):|z|\leq\hbar$ using only translations and
linear canonical transformations $S\in\operatorname*{Sp}(2n,\mathbb{R})$.
Equivalently, $\mathcal{QB}^{2n}$ is an ellipsoid obtained from $B^{2n}%
(\sqrt{\hbar})$ by an affine symplectic transformation. More precisely:

\begin{definition}
Let $S\in\operatorname*{Sp}(2n,\mathbb{R})$ and $z_{0}\in\mathbb{R}_{z}^{2n}$.
Then $\mathcal{QB}^{2n}(z_{0},S)=T(z_{0})SB^{2n}(\sqrt{\hbar})$ where
$T(z_{0})$ is the translation operator $z\longmapsto z+z_{0}$. Equivalently,
it is the set
\[
\mathcal{QB}^{2n}(z_{0},S)=\{z:(S^{-1})^{T}S^{-1}(z-z_{0})^{2}\leq\hbar\}
\]
where we are writing $(S^{-1})^{T}S^{-1}(z-z_{0})^{2}$ for $(S^{-1})^{T}%
S^{-1}(z-z_{0})^{2}$. The set of all quantum blobs in phase space
$\mathbb{R}_{z}^{2n}$ is denoted $\mathcal{QB}(2n,\mathbb{R}).$
\end{definition}

One shows (de Gosson \cite{Birk}, de Gosson and Luef \cite{golu10}) that a
quantum blob $\mathcal{QB}^{2n}(z_{0},S)$ is characterized by the two
following \emph{equivalent} properties:

\begin{itemize}
\item \textit{The intersection of the ellipsoid }$\mathcal{QB}^{2n}(z_{0}%
,S)$\textit{\ with a plane passing through }$z_{0}$\textit{\ and parallel to
any of the plane of canonically conjugate coordinates }$x_{j},p_{j}%
$\textit{\ in }$\mathbb{R}_{z}^{2n}$ \textit{is an ellipse with area }%
$\pi(\sqrt{\hbar})^{2}=\frac{1}{2}h$\textit{; that area is called the
\emph{symplectic capacity }of the quantum blob }$\mathcal{QB}^{2n}(z_{0},S)$
\textit{(we will discuss more in detail this notion in a moment);}

\item \textit{The supremum of the set of all numbers }$\pi R^{2}%
$\textit{\ such that the ball }$B^{2n}(\sqrt{R}):|z|\leq R$\textit{\ can be
embedded into }$\mathcal{QB}^{2n}(z_{0},S)$\textit{\ using canonical
transformations (linear, or not) is }$\pi(\sqrt{\hbar})^{2}$\textit{. Hence no
phase space ball with radius }$R>\sqrt{\hbar}$ \textit{can be
\textquotedblleft squeezed\textquotedblright\ inside }$\mathcal{QB}^{2n}%
(z_{0},S)$\textit{\ using only canonical transformations (Gromov's
non-squeezing theorem \cite{Gromov}, alias the principle of the symplectic
camel).}
\end{itemize}

It turns out (de Gosson \cite{Birk}) that in the first of these conditions one
can replace the plane of conjugate coordinates with any symplectic plane (a
symplectic plane is a two-dimensional subspace of $\mathbb{R}_{z}^{2n}$ on
which the restriction of the symplectic form $\sigma$ is again a symplectic form).

Clearly there is a natural action%
\begin{gather*}
\operatorname*{Sp}(2n,\mathbb{R})\times\mathcal{QB}(2n,\mathbb{R}%
)\longrightarrow\mathcal{QB}(2n,\mathbb{R})\\
(S,\mathcal{QB}^{2n}(z_{0},S))\longmapsto S[\mathcal{QB}^{2n}(z_{0},S)]
\end{gather*}
of symplectic matrices on quantum blobs: for $S^{\prime}\in\operatorname*{Sp}%
(2n,\mathbb{R})$ we have $S^{\prime}T(z_{0})=T(S^{\prime-1}z_{0})S^{\prime}$
and hence
\begin{equation}
S^{\prime}[\mathcal{QB}^{2n}(z_{0},S)]=T(S^{\prime-1}z_{0})S^{\prime}%
SB^{2n}(\sqrt{\hbar})=\mathcal{QB}^{2n}(z_{0},S^{\prime}S). \label{ssp}%
\end{equation}
Conversely:

\begin{proposition}
\label{prop2}Let $G\in\operatorname*{Sp}(2n,\mathbb{R})$ be positive-definite
and symmetric. The set $\{z:G(z-z_{0})^{2}\leq\hbar\}$ is a quantum blob
$\mathcal{QB}^{2n}(z_{0},S)$.
\end{proposition}

\begin{proof}
As a consequence of the symplectic polar decomposition theorem (see e.g. de
Gosson \cite{Birk}) there exists $S\in\operatorname*{Sp}(2n,\mathbb{R})$ such
that $G=(S^{-1})^{T}S^{-1}$ hence the condition $G(z-z_{0})^{2}\leq\hbar$ is
equivalent to $(S^{-1})^{T}S^{-1}(z-z_{0})^{2}\leq\hbar$.
\end{proof}

The symplectic matrix $S$ defining a given quantum blob is not unique; one
shows (see de Gosson \cite{Birk}) that $\mathcal{QB}^{2n}(z_{0}%
,S)=\mathcal{QB}^{2n}(z_{0},S^{\prime})$ if and only if $S^{\prime}=SU$ where
$U$ is a symplectic rotation, i.e. an element of the subgroup
$U(n)=\operatorname*{Sp}(2n,\mathbb{R})\cap O(2n,\mathbb{R})$ of the
symplectic group. This property reflects the invariance of phase space balls
centered at the origin under rotations. A consequence of this fact is that we
have the following topological identification (de Gosson \cite{de05}):%
\[
\mathcal{QB}(2n,\mathbb{R})\equiv\mathbb{R}^{n(n+1)}\times\mathbb{R}%
^{2n}\equiv\mathbb{R}^{n(n+3)}.
\]
Thus, if we view $\mathcal{QB}(2n,\mathbb{R})$ as a \textquotedblleft quantum
phase space\textquotedblright\ its topological dimension $n(n+3)$ is much
larger than that, $2n$, of the classical phase space, even when $n=1$ (in the
latter case $\dim\mathcal{QB}(2,\mathbb{R})=3$, which is easily understood as
follows: one need one parameter to specify the centre of the quantum blob
(which is here an ellipse with area $h/2$), one to specify one of the
principal axes, and another to describe the angle of a principal axe with,
say, the $x$-axis. A similar interpretation applies in higher dimensions.

Let us briefly compare quantum blobs to the usual quantum cells from
statistical mechanics. A quantum cell is typically a phase space cube with
volume $(\sqrt{h})^{2n}=h$. The first obvious remark is that these cells do
not have any symmetry under general symplectic transformations; while such a
transformation preserves volume, a cube will in general be distorted into a
multidimensional polyhedron. But what is more striking is the comparison of
volumes. Since a quantum blob is obtained from the ball $B^{2n}(\sqrt{\hbar})$
by a volume-preserving transformation its volume is given by%
\[
\operatorname*{Vol}\left(  \mathcal{QB}^{2n}(z_{0},S)\right)  =\frac{h^{n}%
}{n!2^{n}}%
\]
and is hence $n!2^{n}$ smaller than that of a quantum cell. For instance, in
the case of the physical three-dimensional configuration space this leads to a
factor of $48$. In the case of a macroscopic system with $n=10^{23}$ this fact
becomes unimaginably large. This is in strong contrast with the fact that the
orthogonal projection of a quantum blob on any plane $x_{j},p_{j}$ of
conjugate coordinates (or, more generally, on any symplectic plane) is an
ellipse with area equal to $\pi\hbar=h/2$.

\section{The Correspondence $\mathcal{G}$ \label{secg}}

Recall that the Wigner transform of a pure state $\Psi$ is given by%
\begin{equation}
W\Psi(z)=\left(  \tfrac{1}{2\pi\hbar}\right)  ^{n}\int e^{-\frac{i}{\hbar
}p\cdot y}\Psi(x+\tfrac{1}{2}y)\Psi^{\ast}(x-\tfrac{1}{2}y)d^{n}y
\label{Wigner}%
\end{equation}
where the star $^{\ast}$ denotes complex conjugation.

The Wigner transform of the fiducial coherent state $\Phi^{\hbar}$ is given
by
\[
W\Phi^{\hbar}(z)=(\pi\hbar)^{-n}e^{-\frac{1}{\hbar}|z|^{2}.}.
\]
More generally \cite{Birk,Littlejohn} the Wigner transform
\begin{equation}
W\Phi_{M}^{\hbar}(z)=\left(  \tfrac{1}{2\pi\hbar}\right)  ^{n}\int
e^{-\frac{i}{\hbar}p\cdot y}\Phi_{M}^{\hbar}(x+\tfrac{1}{2}y)\Phi_{M}^{\hbar
}(x-\tfrac{1}{2}y)^{\ast}d^{n}y \label{wfi}%
\end{equation}
of the squeezed coherent state $\Phi_{M}^{\hbar}=\Phi_{X,Y}^{\hbar}$ is given
by the formula:%
\begin{equation}
W\Phi_{M}^{\hbar}(z)=(\pi\hbar)^{-n}e^{-\frac{1}{\hbar}Gz\cdot z} \label{wg}%
\end{equation}
where $G$ is the real $2n\times2n$ matrix
\begin{equation}
G=%
\begin{pmatrix}
X+YX^{-1}Y & YX^{-1}\\
X^{-1}Y & X^{-1}%
\end{pmatrix}
. \label{g}%
\end{equation}
Notice that $G$ does not contain the parameter $\hbar$. It turns out that $G$
is both positive definite and symplectic; in fact $G=S^{T}S$ where
\begin{equation}
S=%
\begin{pmatrix}
X^{1/2} & 0\\
X^{-1/2}Y & X^{-1/2}%
\end{pmatrix}
\in\operatorname*{Sp}(2n,\mathbb{R}). \label{ass}%
\end{equation}
The same analysis applies to $\Phi_{M,z_{0}}^{\hbar}(z_{0})$. Letting the
translation operator $T(z_{0}):z\longmapsto z+z_{0}$ act on functions on phase
space by the rule $T(z_{0})f(z)=f(z-z_{0})$ and its quantum variant, the
Heisenberg--Weyl operator (\ref{wtw}) we have the translational property
\begin{equation}
W(\widehat{T}^{\hbar}(z_{0})\psi)(z)=T(z_{0})W(\psi)(z) \label{wtw}%
\end{equation}
and hence, in particular%
\begin{equation}
W\Phi_{M,z_{0}}^{\hbar}(z)=(\pi\hbar)^{-n}e^{-\frac{1}{\hbar}G(z-z_{0})^{2}}.
\label{wg0}%
\end{equation}

Let us now state and prove the following essential correspondence result which
\emph{identifies} squeezed coherent states with quantum blobs:

\begin{proposition}
There is a bijective correspondence
\[
\mathcal{G}:\mathcal{CS(}n,\mathbb{R})\longleftrightarrow\mathcal{QB}%
(2n,\mathbb{R})
\]
between coherent states and quantum blobs. That correspondence is defined as
follows: if
\[
W\Phi_{M,z_{0}}^{\hbar}(z)=(\pi\hbar)^{-n}e^{-\frac{1}{\hbar}G(z-z_{0})^{2}}%
\]
then we have
\begin{equation}
\mathcal{G}[\Phi_{M,z_{0}}^{\hbar}]=\{z:G(z-z_{0})^{2}\leq\hbar\}=\mathcal{QB}%
^{2n}(z_{0},S^{-1}) \label{GH}%
\end{equation}
where the symplectic matrix $S$ is given by formula (\ref{ass}) above.
\end{proposition}

\begin{proof}
While the definition of the correspondence $\mathcal{G}$ is straightforward,
it is not immediately clear why it should be bijective. Let us first show that
it is one-to-one. Suppose that $\mathcal{G}[\Phi_{M,z_{0}}^{\hbar
}]=\mathcal{G}[\Phi_{M^{\prime},z_{0}^{\prime}}^{\hbar}]$, that is%
\[
\{z:G(z-z_{0})^{2}\leq\hbar\}=\{z:G^{\prime}(z-z_{0}^{\prime})^{2}\leq
\hbar\}.
\]
We must then have $G=G^{\prime}$ and $z_{0}=z_{0}^{\prime}$ so that
$W\Phi_{M,z_{0}}^{\hbar}(z)=W\Phi_{M^{\prime},z_{0}^{\prime}}^{\hbar}(z)$;
since the Wigner transform of a function $\Psi$ determines uniquely determines
$\Psi$ up to a unimodular factor we have $\Phi_{M^{\prime},z_{0}^{\prime}%
}^{\hbar}=e^{\frac{i}{\hbar}\gamma}\Phi_{M,z_{0}}^{\hbar}$ for some real phase
$\gamma$. Let us next show that $\mathcal{G}$ is onto; this will at the same
time yield a procedure for calculating the inverse of $\mathcal{G}$. Assume
that $\mathcal{QB}^{2n}(0,S^{-1})=S^{-1}B^{2n}(\sqrt{\hbar})$ is a quantum
blob centered at the origin. One can factorize the matrix $S^{-1}$ as follows
(\textquotedblleft pre-Iwasawa factorization\textquotedblright; cf.
\cite{Birk}, \S 2.2, Corollary 2.30):%
\[
S^{-1}=%
\begin{pmatrix}
L & 0\\
Q & L^{-1}%
\end{pmatrix}%
\begin{pmatrix}
A & -B\\
B & A
\end{pmatrix}
\]
where the symmetric matrix $L$ is given by
\begin{equation}
L=(D^{T}D+B^{T}B)^{1/2} \label{l}%
\end{equation}
is symmetric positive definite,%
\begin{equation}
Q=-(C^{T}D+A^{T}B)(D^{T}D+B^{T}B)^{-1/2} \label{q}%
\end{equation}
with $A+iB\in U(n,\mathbb{C})$. The matrix $%
\begin{pmatrix}
A & -B\\
B & A
\end{pmatrix}
$ is thus a symplectic rotation and, as such, leaves any ball centered at the
origin invariant. Setting $X^{1/2}=L$ and\ $Y=X^{1/2}Q$ it follows that we
have
\[
S^{-1}\left[  B^{2n}(\sqrt{\hbar})\right]  =%
\begin{pmatrix}
X^{1/2} & 0\\
X^{-1/2}Y & X^{-1/2}%
\end{pmatrix}
B^{2n}(\sqrt{\hbar});
\]
the quantum blob $\mathcal{QB}^{2n}(0,S^{-1})$ is thus represented by $Gz\cdot
z\leq\hbar$ where $G=S^{T}S$ is of the type (\ref{g}); define now $\Phi
_{M}^{\hbar}=\Phi_{X,Y}^{\hbar}$ by assigning to $X$ and $Y$ the values
$L^{2}$ and $X^{1/2}Q$ found above. The argument generalizes in a
straightforward way to quantum blobs with arbitrary centre.
\end{proof}

In view of the correspondence between squeezed coherent states and quantum
blobs, we can give a phase space picture of formula (\ref{sol1}) for the time
evolution of a squeezed coherent state when the Hamiltonian function is
quadratic. Let us study this deformation in some detail.

We claim that an initial quantum blob becomes after time $t$ a new quantum
blob which is just its image by the classical flow $S_{t}$:

\begin{proposition}
\label{propmotion}After time $t$ the initial quantum blob $\mathcal{QB}%
^{2n}(z_{0},S_{0})$ becomes the quantum blob%
\[
S_{t}[\mathcal{QB}^{2n}(z_{0},S_{0})]=\mathcal{QB}^{2n}(z_{t},S_{0}S_{t}).
\]
Thus, the quantum motion of coherent states induces the classical motion for
the corresponding quantum blob.
\end{proposition}

\begin{proof}
At initial time we are in presence of an initial quantum blob $\mathcal{QB}%
^{2n}(z_{0},S)$, set of all phase space points $z$ such that $G(z-z_{0}%
)(z-z_{0})\leq\hbar$ with $G=(S^{-1})^{T}S^{-1}$. Let us calculate the Wigner
transform%
\begin{equation}
W\Psi(z,t)=\left(  \tfrac{1}{2\pi\hbar}\right)  ^{n}\int e^{-\frac{i}{\hbar
}p\cdot y}\Psi(x+\tfrac{1}{2}y,t)\Psi^{\ast}(x-\tfrac{1}{2}y,t)d^{n}y
\label{wpsi}%
\end{equation}
of the solution $\Psi(z,t)$ of Schr\"{o}dinger's equation (\ref{schrh}). Using
formula (\ref{solpsi})) together with the symplectic covariance of the Wigner
transform (de Gosson \cite{Birk}) we have%
\[
W\Psi(z,t)=W(\widehat{S}_{t}\Phi_{M_{0},z_{0}}^{\hbar})(z)=W\Phi_{M_{0},z_{0}%
}^{\hbar}(S_{t}^{-1}z).
\]
that is, in view of formula (\ref{wg}) giving the Wigner transform of
$\Phi_{M_{0},z_{0}}^{\hbar}$:
\begin{align}
W\Psi(z,t)  &  =(\pi\hbar)^{-n}\exp\left[  -\frac{1}{\hbar}(S_{t}^{-1}%
)^{T}G_{0}(S_{t}^{-1}z-z_{0})^{2}\right]  .\label{wpsit}\\
&  =(\pi\hbar)^{-n}\exp\left[  -\frac{1}{\hbar}(S_{t}^{-1})^{T}G_{0}S_{t}%
^{-1}(z-z_{t})^{2}\right]  .
\end{align}
It follows that the initial quantum blob has become the ellipsoid defined by%
\[
(S_{t}^{-1})^{T}G_{0}S_{t}^{-1}(z-z_{t})^{2}\leq\hbar
\]
which proves our claim.
\end{proof}

\section{Statistical Interpretation of $\mathcal{G}$}

We begin by recalling the notion of symplectic capacity, which was already
mentioned briefly in the beginning of this paper after the definition of
quantum blobs. See Hofer--Zehnder \cite{HZ}, Polterovich \cite{Polter}, or de
Gosson \cite{go09} and de Gosson and Luef \cite{golu10} for a review of this
notion from point of view easily accessible to physicists.

A symplectic capacity on phase space $\mathbb{R}_{z}^{2n}$ assigns to every
subset $\Omega$ of $\mathbb{R}_{z}^{2n}$ a number $c(\Omega)\geq0$, or
$+\infty$. This assignment must obey the following rules:

\begin{description}
\item[(SC1)] If $\Omega\subset\Omega^{\prime}$ then $c(\Omega)\leq
c(\Omega^{\prime})$;

\item[(SC2)] If $f$ is a canonical transformation then $c(f(\Omega
))=c(\Omega)$;

\item[(SC3)] If $\lambda$ is a real number then $c(\lambda\Omega)=\lambda
^{2}c(\Omega)$; here $\lambda\Omega$ is the set of all points $\lambda z$ when
$z\in\Omega$;

\item[(SC4)] We have $c(B^{2n}(R))=\pi R^{2}=c(Z_{j}^{2n}(R))$; here
$B^{2n}(R)$ is the ball $|x|^{2}+|p|^{2}\leq R^{2}$ and $Z_{j}^{2n}(R)$ the
cylinder $x_{j}^{2}+p_{j}^{2}\leq R^{2}$.
\end{description}

There exist infinitely many symplectic capacities, however the construction of
any of them is notoriously difficult (the fact that symplectic capacities
exist is actually equivalent to Gromov's non-squeezing theorem \cite{Gromov}).
However they all agree on phase space ellipsoids. In fact:

\begin{proposition}
\label{propsymp}Let $\mathcal{W}:Mz\cdot z\leq\hbar$ where $M$ is a symmetric
positive definite $2n\times2n$ matrix. We have%
\begin{equation}
c(\mathcal{W)=}\pi\hbar/\lambda_{\max} \label{cw}%
\end{equation}
for every symplectic capacity $c$; here $\lambda_{\max}$ is the largest
symplectic eigenvalue of $M.$
\end{proposition}

The proof of this result is based on a symplectic diagonalisation of $M$; see
de Gosson \cite{Birk}, Hofer--Zehnder \cite{HZ}, Polterovich \cite{Polter},
and the references therein. Recall that the symplectic eigenvalues of $M$ are
defined as follows: the eigenvalues of the matrix $JM$ are of the type $\pm
i\lambda_{j}$ with $\lambda_{j}>0$; the sequence $(\lambda_{1},...,\lambda
_{n})$ is then the symplectic spectrum of $M$ and the $\lambda_{j}$ the
symplectic eigenvalues.

The smallest symplectic capacity is denoted by $c_{\min}$ (\textquotedblleft
Gromov width\textquotedblright): by definition $c_{\min}(\Omega)$ is the
supremum of all numbers $\pi R^{2}$ such that there exists a canonical
transformation such that $f(B^{2n}(R))\subset\Omega$. The fact that $c_{\min}$
really is a symplectic capacity follows from Gromov's \cite{Gromov} symplectic
non-squeezing theorem. For a discussion of Gromov's theorem (and comments)
from the physicist's point of view see de Gosson \cite{go09}, de Gosson and
Luef \cite{golu10}.

Let now $K$ be an arbitrary real symmetric positive-definite matrix of order
$2n$ and define the normalized phase space Gaussian%
\[
W_{K}(z)=(\pi\hbar)^{-n/2}(\det K)^{1/2}e^{-\frac{1}{\hbar}Kz\cdot z}.
\]
When $K=G\in\operatorname*{Sp}(2n,\mathbb{R})$ the Gaussian $W_{K}(z)$ is the
Wigner transform of some squeezed coherent state. Following Littlejohn
\cite{Littlejohn} we define a matrix $\Sigma$ by the relation%
\begin{equation}
\Sigma=\frac{\hbar}{2}K^{-1} \label{covma}%
\end{equation}
hence $W_{K}(z)$ takes the familiar form%
\[
W_{K}(z)=(2\pi)^{-n}(\det\Sigma)^{-1/2}e^{-\frac{1}{2}\Sigma^{-1}z\cdot z}%
\]
suggesting to interpret $\Sigma$ as the covariance matrix of a normal
probability distribution centered at the origin. We will write $\Sigma$ in
block form%
\[
\Sigma=%
\begin{pmatrix}
\Delta(X,X) & \Delta(X,P)\\
\Delta(P,X) & \Delta(P,P)
\end{pmatrix}
\]
where each block has dimension $n\times n$ and $\Delta(P,X)=\Delta(X,P)^{T}$;
we use the notation
\begin{align*}
\Delta(X,X)  &  =(\operatorname*{Cov}(x_{j},x_{k}))_{1\leq j,k\leq n}\\
\Delta(X,P)  &  =(\operatorname*{Cov}(x_{j},p_{k}))_{1\leq j,k\leq n}\\
\Delta(P,P)  &  =(\operatorname*{Cov}(p_{j},p_{k}))_{1\leq j,k\leq n}%
\end{align*}
and set
\[
(\Delta x_{j})^{2}=\operatorname*{Cov}(x_{j},x_{j})\text{ , }(\Delta
p_{j})^{2}=\operatorname*{Cov}(p_{j},p_{j})
\]
for $1\leq j\leq n$. The essential observation we make is:

\begin{proposition}
Consider the phase space ellipsoid $\mathcal{W}:\frac{1}{2}\Sigma^{-1}z\cdot
z\leq1$. The topological condition%
\begin{equation}
c(\mathcal{W)\geq}\frac{1}{2}\hbar\label{capw}%
\end{equation}
implies the Robertson--Schr\"{o}dinger inequalities%
\begin{equation}
(\Delta x_{j})^{2}(\Delta p_{j})^{2}\geq\operatorname*{Cov}(x_{j},p_{j}%
)^{2}+\frac{1}{4}\hbar^{2} \label{RS}%
\end{equation}
for $1\leq j\leq n$ hence, in particular, the Heisenberg uncertainty relations
$\Delta x_{j}\Delta p_{j}\geq\frac{1}{2}\hbar$.
\end{proposition}

The proof of this important result is given in de Gosson \cite{de05,Birk,go09}
(also see de Gosson and Luef \cite{golu10}). It is based on the following
fact, well-known in the quantum optics community: the condition
\begin{equation}
\Sigma+\frac{i\hbar}{2}J\text{ is Hermitian positive semi-definite}
\label{sigma}%
\end{equation}
implies the Robertson--Schr\"{o}dinger inequalities (\ref{RS}) (but it is not
equivalent to it: see de Gosson \cite{go09} for a counterexample). Some
algebra together with a formula giving the symplectic capacity of an
ellipsoid, then shows that conditions (\ref{sigma}) and (\ref{capw}) are
equivalent. Notice that the matrix $\Sigma+\frac{i\hbar}{2}J$ is always
Hermitian since $(\Sigma+\frac{i\hbar}{2}J)^{\ast}=\Sigma-\frac{i\hbar}%
{2}J^{T}$ and $J^{T}=-J$. We mention that symplectic capacities can be used as
well for the study of the more general uncertainty principle related to
non-commutative quantum mechanics as we have shown in de Gosson
\cite{degostat}.

Suppose now that the covariances defined above correspond to some quantum
state $\Psi$ (pure or mixed). The Robertson--Schr\"{o}dinger inequalities
(\ref{RS}) are saturated (i.e. they become equalities) exactly when that state
is a squeezed coherent state $\Phi_{M}^{\hbar}$ where $M=X+iY$ is determined
via the Wigner transform of $\Phi_{M}^{\hbar}$ (cf. (\ref{covma}))
\[
W\Phi_{M}^{\hbar}(z)=(\pi\hbar)^{-n}e^{-\frac{1}{\hbar}Gz\cdot z}\text{ \ ,
\ }G=\frac{\hbar}{2}\Sigma^{-1}.
\]
For instance if $\Phi_{M}^{\hbar}$ is the fiducial coherent state $\Phi
^{\hbar}$ all the covariances vanish and the inequalities (\ref{RS}) reduce to
$\Delta x_{j}\Delta p_{j}=\frac{1}{2}\hbar$ for $1\leq j\leq n$.

\section{Fermi's Function $g_{\mathrm{F}}$ \label{secfe}}

In a largely forgotten paper from 1930 Fermi \cite{Fermi} associates to every
quantum state $\Psi$ a certain hypersurface $g_{\mathrm{F}}(x,p)=0$. Fermi's
paper has recently been rediscovered by Benenti \cite{benenti} and Benenti and
Strini \cite{best}; in particular these authors give a heuristic comparison of
the function $g_{\mathrm{F}}$ and the Wigner transform $W\Psi$. Let us shortly
study the relationship between Fermi's function and the notion of quantum
blob. The starting point is Fermi's observation that the state of a quantum
system may be defined in two different (but equivalent) ways, namely by its
wavefunction $\Psi$ or by measuring a certain physical quantity whose
definition goes as follows. Writing the wavefunction in polar form
$\Psi(x)=R(x)e^{i\Phi(x)/\hslash}$ ($R(x)\geq0$ and $\Phi(x)$ real) one
verifies by a straightforward calculation that $\Psi$ is a solution of the
partial differential equation
\begin{equation}
\widehat{g_{\mathrm{F}}}\Psi=0\label{gfpsi}%
\end{equation}
where
\begin{equation}
\widehat{g_{\mathrm{F}}}=\left(  -i\hbar\nabla_{x}-\nabla_{x}\Phi\right)
^{2}+\hbar^{2}\frac{\nabla_{x}^{2}R}{R}.\label{gf1}%
\end{equation}
The equation (\ref{gfpsi}) seems at first sight to be ad hoc and somewhat
mysterious. However much of the mystery disappears if one remarks that this
equation is obtained by the gauge transform $p\longrightarrow p-\nabla_{x}%
\Phi$ from the trivial equation%
\begin{equation}
\left(  -\hbar^{2}\nabla_{x}^{2}+\hbar^{2}\frac{\nabla_{x}^{2}R}{R}\right)
R=0.\label{trivial}%
\end{equation}

Consider now the Weyl symbol of the operator $\widehat{g_{\mathrm{F}}}$; it is
the real function%
\begin{equation}
g_{\mathrm{F}}(x,p)=\left(  p-\nabla_{x}\Phi\right)  ^{2}+\hbar^{2}%
\frac{\nabla_{x}^{2}R}{R}.\label{gf2}%
\end{equation}
When $\nabla_{x}^{2}R/R<0$ the equation $g_{\mathrm{F}}(x,p)=0$ determines a
hypersurface $\mathcal{H}_{\mathrm{F}}$ in phase space $\mathbb{R}_{z}^{2n}$
which Fermi ultimately \emph{identifies} with the state $\Psi$. Let us examine
the relation between Fermi's Ansatz and the notion of quantum blob we have
introduced in this paper. Let $\Phi_{M}^{\hbar}=\Phi_{X,Y}^{\hbar}$ be the
squeezed coherent state defined by Eqn. (\ref{Gauss1}); we have in this case
$\Phi(x)=-\frac{1}{2}Yx\cdot x$ and $R(x)=e^{-Xx\cdot x/2\hbar}$ hence Fermi's
function is%
\begin{equation}
g_{\mathrm{F}}(x,p)=\left(  p+Yx\right)  ^{2}+X^{2}x\cdot x-\hbar
\operatorname*{Tr}X\label{gf3}%
\end{equation}
where $\operatorname*{Tr}X$ is the trace of the matrix $X$ (note that
$\operatorname*{Tr}X>0$ since $X$ is positive definite). The hypersurface
$\mathcal{H}_{\mathrm{F}}$ is thus the closed hypersurface%
\begin{equation}
M_{\mathrm{F}}z\cdot z=\hbar\text{\ \ with \ }M_{\mathrm{F}}=\frac
{1}{\operatorname*{Tr}X\text{ }}%
\begin{pmatrix}
X^{2}+Y^{2} & Y\\
Y & I
\end{pmatrix}
.\label{mf}%
\end{equation}
Recall now that the Wigner transform of $\Phi_{M}^{\hbar}$ is the function
$W\Phi_{M}^{\hbar}(z)=(\pi\hbar)^{-n}e^{-\frac{1}{\hbar}Gz\cdot z}$ where
(formulas (\ref{g}) and (\ref{ass}))
\begin{equation}
G=%
\begin{pmatrix}
X+YX^{-1}Y & YX^{-1}\\
X^{-1}Y & X^{-1}%
\end{pmatrix}
=S^{T}S\label{gg}%
\end{equation}
and $S$ is the symplectic matrix
\begin{equation}
S=%
\begin{pmatrix}
X^{1/2} & 0\\
X^{-1/2}Y & X^{-1/2}%
\end{pmatrix}
.\label{ess}%
\end{equation}
An immediate calculation shows that the matrices $M_{\mathrm{F}}$ and $G$ are
related by the formula%
\begin{equation}
M_{\mathrm{F}}=\frac{1}{\operatorname*{Tr}X\text{ }}S^{T}%
\begin{pmatrix}
X & 0\\
0 & X
\end{pmatrix}
S.\label{mfs}%
\end{equation}

Let us consider the \textquotedblleft Fermi ellipsoid\textquotedblright%
\ $\mathcal{W}_{\mathrm{F}}:M_{\mathrm{F}}z\cdot z\leq\hbar$ bounded by the
hypersurface $\mathcal{H}_{\mathrm{F}}$.

\begin{proposition}
(i) There exist symplectic coordinates in which the Fermi ellipsoid
$\mathcal{W}_{\mathrm{F}}:M_{\mathrm{F}}z\cdot z\leq\hbar$ is represented by
the inequality
\begin{equation}
Xx\cdot x+Xp\cdot p\leq\hbar\operatorname*{Tr}X\label{ferx}%
\end{equation}
or by the inequality
\begin{equation}
\sum_{j=1}^{N}\lambda_{j}(x_{j}^{2}+p_{j}^{2})\leq\hbar\operatorname*{Tr}%
X\label{ferell}%
\end{equation}
where $\lambda_{1},...,\lambda_{n}$ are the eigenvalues of $X$; 

(ii) We have
\begin{equation}
c(\mathcal{W}_{\mathrm{F}})=\frac{\pi\operatorname*{Tr}X}{\lambda_{\max}}%
\hbar\geq\frac{1}{2}h\label{cwf}%
\end{equation}
where $\lambda_{\max}$ is the largest eigenvalue of $M_{\mathrm{F}}$ and%
\begin{equation}
\frac{1}{2}h\leq c(\mathcal{W}_{\mathrm{F}})\leq\frac{nh}{2}.\label{nh}%
\end{equation}

\end{proposition}

\begin{proof}
(i) In view of (\ref{mfs}) the inequality $M_{\mathrm{F}}z\cdot z\leq\hbar$ is
equivalent to $%
\begin{pmatrix}
X & 0\\
0 & X
\end{pmatrix}
u\cdot u\leq\hbar\operatorname*{Tr}X$ with $u=Sz$. Let $U$ be a rotation in
$\mathbb{R}^{n}$ diagonalising $X$, that is $X=U^{T}DU$ with
$D=\operatorname{diag}(\lambda_{1},...,\lambda_{n})$. Setting $v=%
\begin{pmatrix}
U & 0\\
0 & U
\end{pmatrix}
u$ the inequality $M_{\mathrm{F}}z\cdot z\leq\hbar$ is now equivalent to
(\ref{ferell}) and one concludes by noting that the matrix $R=%
\begin{pmatrix}
U & 0\\
0 & U
\end{pmatrix}
$ is in $U(n)$ (i.e. a symplectic rotation). (ii) Since symplectic capacities
are invariant by symplectic transformations, it suffices to prove formula
(\ref{cwf}) when $\mathcal{W}_{\mathrm{F}}$ is given by Eqn. (\ref{ferx}) or
by Eqn. (\ref{ferell}). In view of Proposition \ref{propsymp} we have
$c(\mathcal{W}_{\mathrm{F}}\mathcal{)=}\pi\hbar/\lambda_{\max}$ and the
equality in (\ref{cwf}) follows noting that the symplectic spectrum of $X$
consists of precisely the eigenvalues of $X$. The inequality $c(\mathcal{W}%
_{\mathrm{F}})\geq\frac{1}{2}h$ is obvious since $\operatorname*{Tr}%
X\geq\lambda_{\max}$ and the inequality $c(\mathcal{W}_{\mathrm{F}})\leq nh/2$
follows from the fact that $\operatorname*{Tr}X\leq n\lambda_{\max}$.
\end{proof}

In view of the double inequality (\ref{nh}) Fermi ellipsoids are not in
general quantum blobs (except for $n=1$). However each of these ellipsoids
contains a quantum blob. To see this it suffices to show that the ellipsoid
defined by (\ref{ferell}) contains the ball $B(\sqrt{\hbar})$ (because the
image of a quantum blob by a linear symplectic transformation is again a
quantum blob). Now, if $(x,p)$ is in $B(\sqrt{\hbar})$ then
\[
\sum_{j=1}^{N}\frac{\lambda_{j}}{\operatorname*{Tr}X}(x_{j}^{2}+p_{j}^{2}%
)\leq\sum_{j=1}^{N}(x_{j}^{2}+p_{j}^{2})\leq\hbar
\]
since $\lambda_{j}/\operatorname*{Tr}X\leq1$ hence our claim. 

Let us discuss the results above on a few simple examples. For the fiducial
coherent state $\Phi^{\hbar}(x)=(\pi\hbar)^{-n/4}e^{-|x|^{2}/2\hbar}$ we have
$X=I$ and $Y=0$ hence the Fermi ellipsoid $\mathcal{W}_{\mathrm{F}}$ is the
disk $|x|^{2}+|p|^{2}\leq n\hbar$ whose symplectic capacity is $n\pi
\hbar=nh/2$. The operator (\ref{gf1}) is here%
\begin{equation}
\widehat{g_{\mathrm{F}}}=-\hbar^{2}\nabla_{x}^{2}+|x|^{2}-n\hbar\label{gf4}%
\end{equation}
and the relation $\widehat{g_{\mathrm{F}}}\Phi^{\hbar}=0$ is hence equivalent
to%
\begin{equation}
\tfrac{1}{2}(-\hbar^{2}\nabla_{x}^{2}+|x|^{2})\Phi^{\hbar}=\tfrac{1}{2}%
n\hbar\Phi^{\hbar}\label{gf5}%
\end{equation}
which simply states the well--known fact that $\Phi^{\hbar}$ is an eigenvector
of the harmonic oscillator Hamiltonian $\widehat{H}=\tfrac{1}{2}(-\hbar
^{2}\nabla_{x}^{2}+|x|^{2})$ corresponding to the first energy level
$E_{0}=\tfrac{1}{2}n\hbar$. One easily verifies that if $\Psi^{\hbar}$ is the
tensor product of $n$ copies of the (unnormalised) Hermite functions
$xe^{-x^{2}/2\hbar}$ then the equation $\widehat{g_{\mathrm{F}}}\Psi^{\hbar
}=0$ is equivalent to
\begin{equation}
\tfrac{1}{2}(-\hbar^{2}\nabla_{x}^{2}+|x|^{2})\Psi^{\hbar}=\tfrac{3}{2}%
n\hbar\Psi^{\hbar}.\label{gf6}%
\end{equation}
The argument may be repeated, and one finds that the Fermi equation
(\ref{gfpsi}) corresponding to a Hermite function, is always equivalent to the
eigenstate equation for the harmonic oscillator corresponding to that function.

The discussion above can be generalised, using metaplectic covariance
properties, to the case of quantum states of operators corresponding to
arbitrary Hamiltonians $H=\frac{1}{2}Mz\cdot z$ where $M$ is symmetric
positive definite (generalised harmonic oscillator). It is certainly
worthwhile studying what happens in more general cases where the quantum
states are no longer Gaussians; see the following discussion.

\section{Concluding Remarks and Perspectives}

Using the correspondence $\mathcal{G}$ defined in Section \ref{secg} we have
sees that quantum blobs exactly correspond to those quantum states which have
minimum uncertainty in the sense of Robertson--Schr\"{o}dinger. This justifies
our claim that quantum blobs represent the smallest regions of phase space
which make sense from a quantum-mechanical perspective. In fact, contrarily to
what is often believed the Heisenberg inequalities and their stronger version,
the Robertson--Schr\"{o}dinger inequalities (\ref{RS}), are not a statement
about the accuracy of our measurement instruments; their derivation assumes on
the contrary \emph{perfect instruments}. The correct interpretation of these
inequalities is the following (see e.g. Peres \cite{Peres}, p.93): if the same
preparation procedure is repeated a large number of times, and is followed by
either by a measurement of $x_{j}$ , or by a measurement of $p_{j}$, the
results obtained have standard deviations $\Delta x_{j}$ and $\Delta p_{j}$
satisfying these inequalities. Such a process thus makes clear the
impossibility of talking about points in phase space having some intrinsic
meaning (cf. Butterfield's paper \cite{Butter} refuting \textquotedblleft
pointillisme\textquotedblright). We note that in \cite{Dragoman} Dragoman uses
the partition of phase space in quantum blobs to propose a new formulation of
quantum mechanics, based on the following postulates:

\begin{axiom}
It is not possible to localize a quantum particle in a phase space regions
smaller that a quantum blob;
\end{axiom}

\begin{axiom}
The phase space extent of a quantum particle is smaller than a quantum blob.
\end{axiom}

These postulates and their implications for quantum physics certainly deserve
to be discussed further.

In a recent paper \cite{Zeno} Hiley and I study a version of the quantum Zeno
paradox for the Bohm trajectory of a sharply located particle modelled by a
Dirac distribution. We showed in this paper that such a recorded quantum
trajectory (in, for instance, a bubble chamber) is just the classical
trajectory predicted by standard Hamiltonian mechanics. It would be both very
interesting and realistic to study this kind of quantum Zeno effect by
replacing the point-like particle by a squeezed coherent state, that is,
equivalently, by a quantum blob. A good starting point could be Hiley
\cite{bafo} where the relationship between the Wigner--Moyal and Bohm
approaches is elucidated; also the connections with the ideas of Hiley and
collaborators in \cite{Hiley1,hica,hicama} could be useful here. We have seen
in Proposition \ref{propmotion} that a quantum blob evolves \emph{classically}
under the action of the linear Hamiltonian flow determined by a quadratic
Hamiltonian. Of course quadratic Hamiltonians are of a very particular type;
the result above remains approximately valid for arbitrary physical
Hamiltonians, and this with an excellent approximation during generically very
large times (Ehrenfest time, as it is called in the theory of quantum
revivals). This observation could allow us to prove, using the correspondence
$\mathcal{G}$, the following conjecture considerably extending the results in
de Gosson and Hiley \cite{Zeno}:

\begin{conjecture}
When we continuously observe the motion of a quantum blob we see its classical
Hamiltonian motion; i.e. an initial quantum blob $\mathcal{QB}^{2n}$ will be
transformed in the set $f_{t}^{H}(\mathcal{QB}^{2n})$ after time $t$; here
$f_{t}^{H}$ is the classical Hamilton flow (Arnol'd \cite{ar89}, Goldstein
\cite{HGoldstein}).
\end{conjecture}

In Section \ref{secfe} we briefly discussed some elementary properties of the
Fermi function $g_{\mathrm{F}}^{\Psi}$ and of the associated Fermi ellipsoid
$\mathcal{W}_{\mathrm{F}}$. The discussion was actually limited to Gaussian
states. We make the following conjecture:

\begin{conjecture}
Let $\Psi$ be a quantum state for which the Fermi equation $g_{\mathrm{F}%
}(x,p)$ defines a hypersurface in phase space bounding a compact set
$\Omega_{\mathrm{F}}$. Then there exists a symplectic capacity $c$ such that
$c(\Omega_{\mathrm{F}})\geq\frac{1}{2}h$ and $\Omega_{\mathrm{F}}$ contains a
quantum blob.
\end{conjecture}

The observant Reader will perhaps have noticed that the equation
$g_{\mathrm{F}}(x,p)=0$ for a system of particles with mass $m$ can be
rewritten%
\[
\frac{1}{2m}\left(  p-\nabla_{x}\Phi\right)  ^{2}+Q=0
\]
if one introduces the quantum potential
\[
Q=-\frac{\hbar^{2}}{2m}\frac{\nabla_{x}^{2}R}{R}%
\]
familiar from the Bohmian approach top quantum mechanics (see Bohm and Hiley
\cite{BoHi}). There thus seems to be a deep connection between this theory and
the phase space approach which certainly deserves to be elucidated and extended.

I am sure that Basil will be excited by these possibilities, and I look
forward writing new papers with him about the truly fascinating topic of
quantum phase space!\bigskip

\ \ \ \ \ \ \ \ \ \ \ \ \ \ \ \ \ \ \ \ \ \ \ \ \ \ \ \ \ \ \ \ \ \ \ \ \ \ \ \ \ \ \ \ \ \ \ \ \ \ \ \ \ \ \ \ \ \ \ \ Happy
birthday, Basil!

\end{document}